\documentclass[fleqn,usenatbib]{mnras}

\usepackage{newtxtext,newtxmath}

\usepackage[T1]{fontenc}
\usepackage{wrapfig}
\DeclareRobustCommand{\VAN}[3]{#2}
\let\VANthebibliography\thebibliography
\def\thebibliography{\DeclareRobustCommand{\VAN}[3]{##3}\VANthebibliography}

\usepackage{graphicx}	
\usepackage{amsmath}	
\usepackage{csvsimple}
\usepackage{soul}
 \usepackage[
 labelfont=bf,
 singlelinecheck=false 
 ]{caption}
\usepackage[ruled,vlined]{algorithm2e}
\usepackage[dvipsnames]{xcolor}
\usepackage{capt-of}

\newcommand{\prdLetters}{Phys. Rev. Lett.}

\title[Turbulent Heating in Blazars]{Balancing Turbulent Heating with Radiative Cooling in Blazars}

\author[Zach et al.]{
Zachary Davis,$^{1}$\thanks{E-mail: zkd@purdue.edu}
Jes\'{u}s~M. Rueda-Becerril$^{2}$
and Dimitrios Giannios$^{1}$
\\
$^{1}$Department of Physics, Purdue University, 525 Northwestern Avenue, West Lafayette, IN 47907, USA\\
$^{2}$Center for Computational Relativity and Gravitation, Rochester Institute of Technology, 85 Lomb Memorial Drive, Rochester, NY 14623, USA
}

\date{Accepted XXX. Received YYY; in original form ZZZ}

\pubyear{2022}

\begin{document}
\label{firstpage}
\pagerange{\pageref{firstpage}--\pageref{lastpage}}
\maketitle

\begin{abstract}
Recently, particle in cell (PIC) simulations have shown that relativistic turbulence in collisionless plasmas can result in an equilibrium particle distribution function where turbulent heating is balanced by radiative cooling of electrons. Strongly magnetized plasmas are characterized by higher energy peaks and broader particle distributions. In relativistically moving astrophysical jets, it is believed that the flow is launched Poynting flux dominated and that the resulting magnetic instabilities may create a turbulent environment inside the jet, i.e., the regime of relativistic turbulence. In this paper, we extend previous PIC simulation results to larger values of plasma magnetization by linearly extrapolating the diffusion and advection coefficients relevant for the turbulent plasmas under consideration. We use these results to build a single zone turbulent jet model that is based on the global parameters of blazar emission region, and consistently calculate the particle distribution and resulting synchrotron and inverse Compton emission spectra. We then test our model by comparing its predictions with the broad-band quiescent emission spectra of a dozen blazars. Our results show good agreement with observations of low-synchrotron peaked (LSP) sources and find that LSPs are moderately Poynting flux dominated with magnetization $1\lesssim \sigma \lesssim 5$, have bulk Lorentz factor $\Gamma\sim 10-30$, and that the turbulent region  is located at the edge, or just beyond, the broad line region (BLR). The turbulence is found to be driven at an area comparable to the jet cross section.
\end{abstract}




\section{Introduction}
\label{sec:intro}
In many astrophysical plasma flows, including those in supernova remnants, pulsar wind nebulae (PWN) or active galactic nuclei (AGN), a broadband emission spectrum of electromagnetic radiation is often observed. AGNs, with a jet closely aligned to our line of sight, are referred to as blazars \citep{urry_1995}. Blazars have a characteristic double peaked spectral energy distribution (SED). The first peak is attributed to synchrotron emission by ultrarelativistic leptons, and the second is likely to be result of inverse Compton (IC) scattering off the same particles \citep{Ghisellini1998}. Blazars also exhibit intense flaring on short timescales followed by quiescent intervals. Both the quiescent and flaring blazar SEDs are routinely explained by an extended, non-thermal, lepton distribution that is usually modelled with a power-law or broken power law \citep{Ghisellini1998}.

For the inferred non-thermal tails of the particle distribution to develop, an efficient particle acceleration mechanism needs to be in place, i.e., where the acceleration time scale is shorter than, or of the order of, the  variability timescale in the emission region. The variability in the emission, especially for fast-evolving flares, puts strong constraints on the acceleration timescales and the size of the emitting regions.  There is an active debate on the particle acceleration mechanisms responsible  for blazar flaring where shocks inside the jet flow \citep[e.g.,][]{Maddalena2001,Bottcher2010,Mimica2012} or magnetic instabilities that result in magnetic reconnection in the jet \citep[e.g.,][]{Giannios2013} are commonly invoked. Regardless of the mechanism that powers the flaring events, at their non-linear stages, the dissipative mechanisms can be expected to drive turbulence within the jet flow \citep{Marscher2016,Baring2016,Comisso2019}.
Turbulence in a strongly magnetized plasma (with magnetic energy density exceeding the plasma enthalpy density; also referred to as \textit{relativistic turbulence}) has long been suspected to be an  acceleration process for relativistic particles \citep{schlickeiser1989}. With recent MHD simulations suggesting that jets are launched as magnetically dominated plasma flows \citep{2007Komissarov,2009Tchekhovskoy,2017Duran}, we may expect relativistic turbulence to drive part of the emission inside these outflows. In this work, we focus on a scenario where the more efficient particle acceleration processes operate at the onset of the jet instabilities and may power blazar flares while the resultant turbulence may be able to drive the quiescent and slow-evolving emission observed in blazars.

Our understanding of relativistic turbulence has substantially advanced recently thanks to  particle-in-cell (PIC) simulations that explore particle acceleration in highly magnetized, turbulent plasmas \citep{Comisso2018,Zhdankin2019}. These simulations have shown that particles undergo an initial rapid acceleration phase from the current sheets created by the turbulence. After this, Alfvén wave scattering, a second order Fermi process \citep{Fermi1949}, begins to dominate the acceleration and produces a non-thermal tail in the particle distribution \citep{Comisso2018}.

In the absence of substantial particle cooling, PIC simulations find that relativistic turbulence energizes particles to the system size-limited energy \citep{Zhdankin2017}. Inside the blazar emission region, however, we expect radiative losses to effectively cool the plasma resulting in a steady state particle distribution as seen in \citet{Uzdensky2018}. The effects of radiative losses are particularly important to understand when studying the particle distribution in relativistic jets where radiative cooling time scales are short. Currently, there are only a handful of PIC simulations that have studied  relativistic turbulence that have also included radiative cooling in the simulation (\emph{radiative relativistic turbulence}). In particular, the results reported by \citet{Zhdankin2020} confirm the analytical results in \citet{Uzdensky2018}, concluding that steady states can be formed in a turbulent radiative plasma. In the same manner, \citet{Comisso2019} show in their simulations that non-thermal tails develop in the particle distribution. Furthermore, the hard tail diffusion seems well described by Alfvén wave scattering theories \citep[see][]{schlickeiser1989}. The previous works in relativistic turbulence mentioned above, though mostly for lower magnetization $\sigma \lesssim 10$, lends insight to the turbulent plasma properties inside of a blazar jet and similar blazar jet like environments. Though PIC simulations have greatly extended our understanding of relativistic plasmas, due to their computational cost, have only studied a small range of plasma magnetization and usually only include a few, if any, radiation mechanisms operating in jets.

In this work, we use latest PIC findings for particle acceleration in relativistic turbulence, generalize the description of the acceleration terms for arbitrary magnetization $\sigma$, incorporate radiative cooling and calculate the equilibrium particle distribution. We then proceed to build a simple single-zone model for the bulk properties of the turbulent region as expected in blazars and apply the model to a dozen sources with broad-band SED spectra. The target is two-fold: (i) evaluate the feasibility of the model in accounting for the quiescent blazar SED and (ii) extracting important properties of the blazar zone such as bulk Lorentz factor, magnetization and distance of the blazar zone from the central engine. This paper is organized as follows: Sec.~\ref{sec:model} outlines our turbulent model. In Sec.~\ref{sec:results} we describe the initial setup and operation of the fitting algorithm used to test our model,as well as the best fit results. In Sec.~\ref{sec:discussion}, we further discuss our results in the context of blazar jet modeling. Finally, in Sec.~\ref{sec:conclusion} we present the conclusion from our findings.

\section{Equilibrium particle distribution from turbulent Acceleration and radiative cooling}
\label{sec:model}

In the present section we build the model that describes the particle distribution of the fluid in the blazar emission region. The particle distribution is found as an equilibrium between turbulent acceleration of particles and the radiative cooling mechanisms operating in these sources.

Turbulence is generated by large scale fluctuations that create a driving current at the boundary of the turbulent region, where energy cascades down to smaller scales via Alfvén waves \citep[see][]{Goldreich1995}. The energy injected into the system through this process is a fraction of the stochastic magnetic energy that propagates the waves. Following \citet{Zhdankin2020}, we will consider turbulence in the strong regime, where the fluctuations in magnetic field strength are comparable to the underlying background magnetic field, i.e., $\delta B_{ \rm rms} \approx B_{0}$. The energy stored in the turbulent magnetic field will be dissipated into the particles over an Alfvén crossing time $\tau_{ \rm a} \equiv R_{ \rm T} / v_{ \rm A}$, where $R_{ \rm T}$ is the scale of the turbulence and $v_{ \rm A} \equiv c\sqrt{\frac{\sigma}{\sigma+1}}$ is the Alfvén speed. With this, we parameterize the mean injected power as
\begin{equation}
\label{eq:turbulent_injection_energy}
\langle \dot{E}_{ \rm \rm inj} \rangle=\eta_{ \rm \rm inj} \frac{B_{0}'^{2}}{8 \pi n_{0}\tau_{ \rm a}},
\end{equation}
where $\eta_{ \rm inj}$ is the fraction of turbulent magnetic energy deposited into the electrons and $n_{0}$ is the particle number density of the electrons. The injected energy \eqref{eq:turbulent_injection_energy}, will heat the fluid until it escapes the turbulent area, or radiative losses balance the heating and create a steady state \citep{Uzdensky2018}.

\subsection{The Particle Acceleration Model}

The evolution of a particle energy distribution is described by the kinetic equation
\begin{equation}
\label{eq:FP_jesus}
\begin{split}
  \dfrac{\partial n(\gamma, t)}{\partial t} & = \dfrac{1}{2} \dfrac{\partial^2}{\partial\gamma^2} \left[ D(\gamma, t) n(\gamma, t) \right] + \dfrac{\partial}{\partial\gamma} \left[ \dot{\gamma}(\gamma, t) n(\gamma, t) \right] + \\
  & + Q(\gamma, t) - \dfrac{n(\gamma, t)}{t_{ \rm esc}},
\end{split}
\end{equation}
also known as the Fokker-Planck equation, where $Q(\gamma, t)$ is the particle injection rate, $D(\gamma, t)$ the particle diffusion coefficient, $n(\gamma, t)$ the differential particle distribution function, $\dot{\gamma}(\gamma, t)$ the energy loss rate, $\gamma$ the particle Lorentz factor, $t$ the time variable and $t_{ \rm esc}$ is the particle escape time, i.e., the average time it takes for a particle to leave the system. This paper works under the assumption that the particle distribution starts and evolves isotropically. Though it should be noted that anisotropies in the particle distribution have been found at high energies \citep{Comisso2019}. Since we are interested in studying the fluid under a steady state, where the turbulent region experiences particle injection at the same rate as particles escape, we set the injection rate to be equal to the escape rate such that they cancel out everywhere. For the same effect on the distributions, we consider the case where $Q(\gamma, t) = 0$, and $t_{ \rm esc} \rightarrow \infty$.

Since in this work we are using the code Paramo \citep{Paramo2019} to solve the Fokker-Planck equation and calculate the emissivity, which uses the Fokker-Planck in the same form as equation \ref{eq:FP_jesus}, we need to find equivalent diffusion and cooling terms to the ones described in \citet{Zhdankin2020}, which choose to introduce the Fokker-Planck in the form
\begin{equation}
\label{eq:FPZhdankin2020}
\dfrac{\partial n(\gamma, t)}{\partial t}=\dfrac{\partial}{\partial \gamma}\left(\gamma^{2} D_{ \rm p p} \dfrac{\partial}{\partial\gamma}\left(\frac{n(\gamma, t)}{\gamma^{2}}\right)\right)-\dfrac{\partial}{\partial \gamma}\left(A_{ \rm p} n(\gamma, t)-\frac{\gamma^{2}}{\gamma_{0} \tau_{ \rm c}} n(\gamma, t)\right).
\end{equation}
The last two expressions can be made equivalent by making the substitution,
\begin{equation}
\label{eq:Diffusioncoeff}
D(\gamma,t)=2D_{ \rm pp},
\end{equation}
for the diffusion and,
\begin{equation}
\label{eq:gammadot}
\dot{\gamma} = -\left(A_{ \rm p}+\frac{1}{\gamma^{2}} \partial_{ \rm \gamma}\left(\gamma^{2} D_{ \rm pp}\right)-\frac{\gamma^{2}}{\gamma_{0} \tau_{ \rm c}}\right),
\end{equation}
for the energy loss term. Where the term $\frac{\gamma^{2}}{\gamma_{0} \tau_{ \rm c}}$, is the radiative cooling term discussed in section \ref{sec:cooling}, $\gamma_{0}$ is the mean Lorentz factor, and $\tau_{\rm c}$ is the cooling time scale.

Following the work done by \citet{Zhdankin2020}, we model the diffusion coefficient quadratically in momentum and the advection coefficient linearly in momentum,
\begin{equation}
\label{eq:difandadv}
\begin{array}{c}
A_{ \rm p} = (\Gamma_{ \rm h} \gamma_{0}+\Gamma_{ \rm a} \gamma )/\tau_{ \rm c}, \\
D_{ \rm pp} = (\Gamma_{0} \gamma_{0}^{2}+ \Gamma_{ \rm 2} \gamma^{2})/\tau_{ \rm c}.
\end{array}
\end{equation}
Here $\gamma_{0}$ represents the mean Lorentz factor of the particle distribution, $\tau_{ \rm c}$ represents the cooling time discussed in section \ref{sec:cooling} and $\Gamma_{ \rm i}$ are constants that are discussed in section \ref{sec:diffusion}

The initial conditions for the particle energy distribution is setup with a Maxwell–Jüttner distribution profile \citep{maxwell-juttner1911},
\begin{equation}
f(\gamma)=\frac{\gamma^2 \beta}{\Theta K_{ \rm 2}(1/\Theta)}\exp{\frac{-\gamma}{\Theta}}, 
    \label{eq:MJdistro}
\end{equation}
with $\Theta = KT/m_{ \rm e}c^{2}$, K is the Boltzmann constant, and $\Theta$ is related to the mean Lorentz factor by $\Theta = \gamma_{0}/3$. Since we are working to a steady state, the initial injection temperature will have little effect on the final distribution \citep{Zhdankin2020}. Thus, the initial distribution is given a temperature very close to $\gamma_{0}/3$. We then allow the distribution to evolve for $t=\tau_{ \rm c}$ at which point, the steady state has been reached.

\subsubsection{Diffusion}
\label{sec:diffusion}

To introduce the latest findings on turbulent particle acceleration from PIC simulations in our model, we use data from \citet{Zhdankin2020}. The data contains the particle distribution for different values of magnetization ranging from $\sigma \sim 0.04-12$. For a given simulation, we time average the distribution after a steady state has been reached. The resultant time averaged distributions are then fitted to the Fokker–Planck steady state eq \eqref{eq:fpss} \citep{Zhdankin2020} using a Markov Chain Monte Carlo (MCMC) method, 
\begin{align}
  f_{ \rm ss}(\gamma) \approx & k \left(\frac{\gamma}{\gamma_{0}}\right)^{2} \left(1+(\frac{\gamma}{\gamma_{0}})^{2}\right)^{\Gamma_{ \rm a} / 2 \Gamma_{ \rm 2}} \nonumber \\
   & \exp \left(-\frac{\gamma}{\gamma_{0}\Gamma_{ \rm 2}}+\frac{\Gamma_{ \rm h}+1}{\Gamma_{ \rm 2}} \tan ^{-1}\left( \frac{\gamma}{\gamma_{0}}\right)\right). \label{eq:fpss}
\end{align}

To maintain consistency with works like \citet{Comisso2019} and \citet{Wong_2020}, where the diffusion coefficient is expected to scale linearly with the magnetization $D_{ \rm \gamma} \sim 0.1 \sigma\left(\frac{c}{l}\right) \gamma^{2}$ (here $l$ is the sytem size), we model the results for $\Gamma_{ \rm i}$ linearly with magnetization, arriving at
\begin{align}
  \Gamma_0 & = \Gamma_2=0.05\sigma + 2.09, \nonumber, \\
  \Gamma_a & = 0.124\sigma - 9.5, \label{eq:gammas} \\
  \Gamma_h & = -0.42\sigma - 2.46. \nonumber
\end{align}

\begin{figure}
    \centering
    \includegraphics[width=0.48\textwidth]{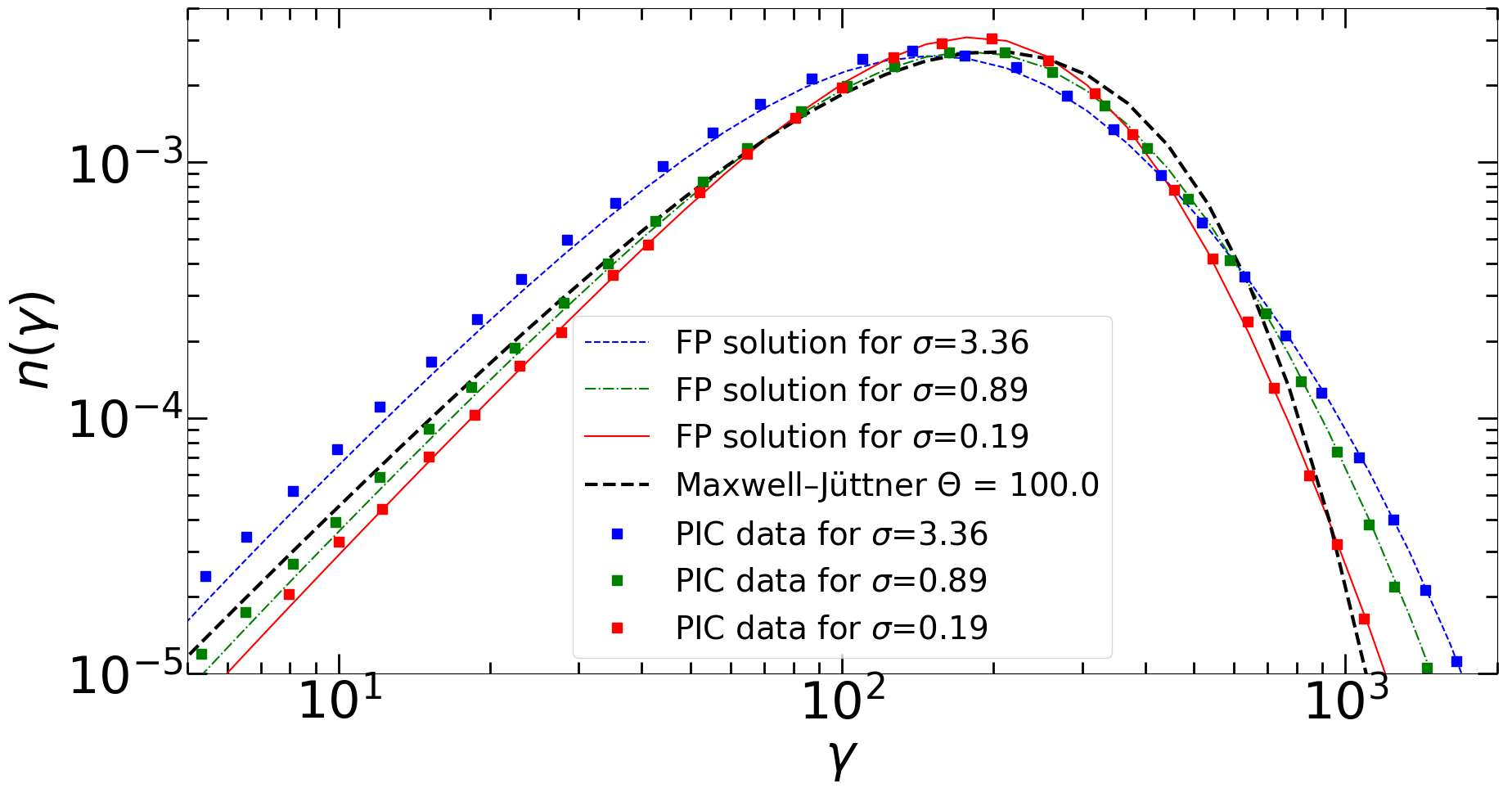}
    \caption{Comparison of the particle distribution found in the PIC simulations of \citet{Zhdankin2020} and our FP solver. The lines are FP solutions created using Paramo \citep{Paramo2019} with parameters to match the setup in \citet{Zhdankin2020}. The points are the steady states found in \citet{Zhdankin2020} figure 17. Red, green, blue curves show the $\sigma =$ 0.19, 0.89, 3.36 cases, respectively. The black dashed curve is the Maxwell–Jüttner distribution for $\Theta = 100$.}
    \label{fig:pd}
\end{figure}

\subsection{Blazar Emission Region}

In this paper, the jet composition is an ultra relativistic electron-ion plasma with cold ions that dominate the plasma's internal energy. This results in a magnetization given by\footnote{Variables in the comoving frame of the plasma will be referenced with a prime symbol ($'$). Non-primed variable are assumed to be in the black-hole rest frame, unless explicitly stated.},
\begin{equation}\label{eq:magnetization}
  \sigma = \frac{2u'_{ \rm B}}{n_{ \rm p} m_{ \rm p}c^{2}},
\end{equation}
where the ion particle density in the comoving frame $n_p$ is expected to be in equal partition with the lepton pair particle density $n_e$ i.e $n_e = n_p = n_{0}$. The magnetic field is assumed to be strongly turbulent so that $B^{'2}_{ \rm rms} = B^{'2}_{0} + \delta B_{ \rm rms}^{'2} = 2B^{'2}_{0}$ is true. Here $B'_{0}$ is related to the jet luminosity by,
\begin{equation}\label{eq:magfromjet}
  B'_{0} = \sqrt{\frac{L_{ \rm j}4\pi}{\Omega_{ \rm j}R^{2}_{ \rm j}\Gamma^{2}_{ \rm j} c}},
\end{equation}
where $\Omega_{ \rm j} = 2\pi (1-\cos(1/\Gamma_{ \rm j})) \approx \pi/\Gamma_{ \rm j}^{2}$ is the jet's solid angle (the approximation is not used in this paper), $L_{ \rm j}$ is the jets luminosity, $R_{ \rm j}$ is the distance from the black hole, $\Gamma_{ \rm j}$ is the jets bulk Lorentz factor, and we assume the jet opening angle is $1/\Gamma_{j}$. Further, the turbulent scale $R_{ \rm T}$, is fraction of the jets cross section, 
\begin{equation}\label{eq:turbulentscale}
  R_{ \rm T} = R_{ \rm TM} \frac{R_{ \rm j}}{\Gamma_{ \rm j}},
\end{equation}
where $R_{ \rm TM}$ is the fore mentioned fraction.

The particles accelerated by the jet are subject to radiation fields produced elsewhere in the blazar environment. Here, we assume the material is exposed to a radiation field from within the broad line region (BLR). The BLR radiation field is assumed to be isotropic and monochromatic with frequency $\nu_{0} = 10^{15}$ Hz and in the comoving frame, $\nu'_{0} = \Gamma_{ \rm j} \nu_{0}$. We parameterize the BLR radiation in lab frame using \citet{Ghisellini_2013},
\begin{equation}\label{eq:blrenergydensitypre}
 u_{ \rm ph}= \frac{L_{ \rm BLR}}{4 \pi c R^{2}_{ \rm BLR}},
\end{equation}
where $L_{ \rm BLR} \approx \eta_{ \rm ph} L_{ \rm disk}$ and $R_{ \rm BLR}\approx 10^{17} \sqrt{\frac{L_{ \rm disk}}{10^{45}}}$ cm \citep{Ghisellini_2013}. Further, our jet luminosity $L_{ \rm j}$ is modeled here to be directly proportional to the accretion power, i.e.,  $L_{ \rm j} = \eta_{ \rm j} \dot{M} c^{2}$. Similarly, we model the disk luminosity as directly proportional to the jet luminosity such that, $L_{ \rm d} = \frac{\eta_{ \rm d}}{\eta_{ \rm j}} L_{ \rm j}$. For the coefficients $\eta_d$ and $\eta_{ \rm j}$ we refer to \citet{Rueda_Becerril_2021} where $\eta_{ \rm j} \approx 1$ and $\eta_{ \rm d} \approx 0.1$,
\begin{equation}
\label{eq:blrenergydensity}
u_{ \rm ph}\approx \eta_{ \rm ph} 0.26 \text{  erg cm$^{-3}$},
\end{equation}

where $\eta_{ \rm ph}$ quantifies the amount of energy from the BLR photons that enter the emission region and $u'_{ \rm ph}=\Gamma^{2}_{ \rm j}(1+\beta^{2}/3) u_{ \rm ph}$ is its value in the comoving frame \citep{Dermer2009}. This description suffices so long as the emission region is within the BLR. Outside of the BLR the photon density drops precipitously. To model this we follow \citet{Nalewajko2014}, where inside the BLR we use the expression in \ref{eq:blrenergydensity}. However, outside the BLR, the photon energy density drops with the cube of distance,
\begin{equation}\label{eq:blrenergydensity2}
 u_{ \rm ph} = \eta_{ \rm ph}  0.26\text{ erg cm$^{-3}$} \times \left\{
  \begin{array}{ll}
    1 & \text {for } \quad R_{ \rm j} \leqslant R_{ \rm BLR} \\
    \left(\frac{R_{ \rm j}}{R_{ \rm BLR}}\right)^{-3} & \mbox{for} \quad R_{ \rm j} > R_{ \rm BLR}
  \end{array}
  \right.,
\end{equation}
with
\begin{equation}\label{eq:rblr}
  R_{ \rm BLR} = 10^{17} \sqrt{\frac{L_{ \rm d}}{10^{45}\text{erg s$^{-1}$}}} \text{cm}  =  10^{17} \sqrt{\frac{\eta_{ \rm d}L_{ \rm j}}{\eta_{ \rm j}10^{45}\text{erg s$^{-1}$}}}\text{cm}.
\end{equation}

\subsection{Cooling \bf{and Emission}}
\label{sec:cooling}
So far, we have discussed how the particles are accelerated in the turbulent region. The particle acceleration is eventually balanced by radiative losses. Relativistic leptons in the blazar environment suffer from synchrotron and Compton losses. Here, we consider both the synchrotron self Compton (SSC) and external inverse Compton (EIC) processes but limit our discussion to scattering in the Thomson limit (i.e., relativistic corrections to the electron scattering cross section are ignored). For a relativistic plasma, the power lost via synchrotron and EIC can be found at \citep{Rybicki:1979},
\begin{subequations}\label{eq:synchandeicpower}
\begin{align}
  \dot{\mathcal{E}}_{ \rm syn} & = \frac{4}{3}\sigma_{ \rm T}c u'_{ \rm B}\gamma^{2},\\
  \dot{\mathcal{E}}_{ \rm eic} & = \frac{4}{3}\sigma_{ \rm T}c u'_{ \rm ph}\gamma^{2},
\end{align}
\end{subequations}
where $u'_{ \rm B}$ is the magnetic energy density and is given by, 
\begin{equation}
    \label{eq:magnetic_energy_density}
    u'_{ \rm B} = \frac{B^{'2}_{ \rm rms}}{8\pi} = \frac{B_0^{'2} + \delta B_{ \rm rms}^{'2}}{8\pi} = \frac{2B^{'2}_{0}}{8\pi}.
\end{equation}
The plasma cooling is assumed to be dominated by synchrotron and EIC but the SSC component is added for completeness,
\begin{equation}\label{eq:totalrad}
  \dot{\mathcal{E}}_{ \rm rad} = \dot{\mathcal{E}}_{ \rm eic} + \dot{\mathcal{E}}_{ \rm syn} + \dot{\mathcal{E}}_{ \rm ssc}.
\end{equation}
For the SSC cooling we follow \citet{Schlickeiser2009} (derivation in appendix \ref{sec:ssc_deriv}),
\begin{equation}
    \label{eq:sscoft}
    \begin{array}{c}
    \dot{\mathcal{E}}_{ \rm ssc} \approx \frac{3\pi\sigma_{ \rm T} c_{ \rm 1} q_{0} \epsilon_{0}^2 R_{ \rm T}}{2 h^{2}} \gamma^{2} \int_{0}^{\infty} d\gamma \gamma^{2}n(\gamma, t)
    \\
    = \frac{3\pi\sigma_{ \rm T} c_{ \rm 1} q_{0} \epsilon_{0}^2 R_{ \rm T} n_{0} }{2 h^{2}} \gamma^{2} \langle \gamma(t)^{2} \rangle .
    \end{array}
\end{equation}
The constant $c_{ \rm 1}$ is found in \citet{schlickeiser1989},
\begin{equation}
\label{eq:c1}
c_{ \rm 1}=\int_{0}^{\infty} \mathrm{d} x x C S(x)=\frac{32}{81} \sqrt{3}=0.684,
\end{equation}
where $\epsilon_{0}$ and $q_{0}$ are given by,
\begin{equation}
    \label{eq:epandq}
    \begin{array}{c}
    \epsilon_{0} \approx \frac{heB'_{ \rm rms}}{2\pi m_{ \rm e} c},
    \\
    q_{0} = \frac{4\pi e^{2}}{\sqrt{3}c}.
    \end{array}
\end{equation}
Since we are interested in the quiescent emission from a turbulent plasma that will reach a steady state, we assume that the SSC has maximum effect once the steady state is reached. With this assumption equation \eqref{eq:sscoft} becomes, 
\begin{equation}
    \label{eq:sscoftc}
    \dot{\mathcal{E}}_{ \rm ssc} \approx \frac{3\pi\sigma_{ \rm T} c_{ \rm 1} q_{0} \epsilon_{0}^2 R_{ \rm T} n_{0} }{2 h^{2}} \gamma^{2} \langle \gamma(\tau_{ \rm c})^{2}\rangle.
\end{equation}
To find the equilibrium energy ($\gamma_{0} m_{ \rm e} c^2$), we balance the injected power with the radiated power,
\begin{equation}
    \label{eq:balance_power}
    \langle \dot{\mathcal{E}}_{ \rm inj} \rangle =  \langle \dot{\mathcal{E}}_{ \rm rad} \rangle.
\end{equation}
Where $ \langle \dot{\mathcal{E}}_{ \rm rad} \rangle$ is the summation of all radiative losses:
\begin{equation}
    \label{eq:averageradiationpower}
    \begin{array}{c}
    \langle\dot{\mathcal{E}}_{ \rm syn}\rangle = \frac{4}{3}\sigma_{ \rm T}c u'_{ \rm B}\langle\gamma^{2}\rangle,
    \\
    \langle\dot{\mathcal{E}}_{ \rm eic}\rangle =  \frac{4}{3}\sigma_{ \rm T}c u'_{ \rm ph}\langle\gamma^{2}\rangle,
    \\
    \langle\dot{\mathcal{E}}_{ \rm ssc}\rangle \approx \frac{3\pi\sigma_{ \rm T} c_{ \rm 1} q_{0} \epsilon_{0}^2 R_{ \rm T} n_{0} }{2 h^{2}} (\langle \gamma^{2}\rangle)^{2}.
    \end{array}
\end{equation}
Since the steady states formed in PIC simulation \citet{Zhdankin2020} can be approximated by a Maxwell–Jüttner distribution, we use the mean squared relation $\langle \gamma^{2}\rangle = \frac{4}{3} \langle \gamma \rangle^{2} = \frac{4}{3} \gamma_{0}^{2}$. With this equation \eqref{eq:averageradiationpower} becomes:
\begin{equation}
    \label{eq:gam0radiationpower}
    \begin{array}{c}
    \langle\dot{\mathcal{E}}_{ \rm syn}\rangle = \frac{16}{9}\sigma_{ \rm T}c u'_{ \rm B}\gamma_{0}^{2},
    \\
    \langle\dot{\mathcal{E}}_{ \rm eic}\rangle =  \frac{16}{9}\sigma_{ \rm T}c u'_{ \rm ph}\gamma_{0}^{2},
    \\
    \langle\dot{\mathcal{E}}_{ \rm ssc}\rangle \approx \frac{ 64\pi^{2} \sqrt{3}\sigma_{ \rm T} c_{ \rm 1} e^{4}  R_{ \rm T} n_{0} u'_{ \rm B} }{ m^{2} c^{3}}\gamma_{0}^{4}
    \\
    = A_{0} R_{ \rm T} n_{ \rm e} u'_{ \rm B} \gamma_{0}^{4}.
    \end{array}
\end{equation}
Plugging \eqref{eq:gam0radiationpower} into \eqref{eq:balance_power} we get a quadratic for $\gamma_{0}^{2}$,
\begin{equation}
    \label{eq:quadraticgam02}
    \begin{array}{c}
    A_{0} R_{ \rm T} n_{ \rm e} \gamma_{0}^{4} +  \frac{16}{9}\sigma_{ \rm T}c (1 + \frac{u'_{ \rm ph}}{u'_{ \rm B}})\gamma_{0}^{2} -  \frac{\eta_{ \rm inj} v_{ \rm a}}{2n_{0}R_{ \rm T}} =0,
    \\
    a_{ \rm 1}\gamma_{0}^{4} + b_{ \rm 1}\gamma_{0}^{2} + c_{ \rm 2} = 0.
    \end{array}
\end{equation}
Solving for $\gamma_{0}$ leads to,
\begin{equation}
    \label{eq:gam0}
    \gamma_{0} = \left(\frac{\left(b_{ \rm 1}^{2} - 4a_{ \rm 1}c_{ \rm 2} \right)^{\frac{1}{2}}  - b_{ \rm 1}}{2a_{ \rm 1}} \right)^{\frac{1}{2}}.
\end{equation}
Solving for $\frac{1}{\tau_{ \rm c}} = \frac{\langle\dot{\gamma}\rangle}{\gamma_{0}}$,
\begin{equation}
\label{eq:tauc}
    \tau_{ \rm c} = \left(\frac{16 \sigma_{ \rm T} c (u'_{ \rm B} + u'_{\rm ph})}{9 m_{ \rm e} c^{2}}\gamma_{0}  + \frac{A_{0} n_{0} u'_{ \rm B} R_{ \rm T}}{ m_{ \rm e} c^{2}}\gamma_{0}^{3}  \right)^{-1},
\end{equation}
where the $\tau_{\rm c} $ is the cooling timescale of an electron with a Lorentz factor $\gamma_{0}$.

The last expressions for $\gamma_{0}$ and $\tau_{ \rm c}$ are simplified when EIC or synchrotron are the dominant cooling processes. When SSC is negligible the cooling is dominated by EIC and synchrotron giving,
\begin{equation}
    \label{eq:gam0_limit_nossc}
    \begin{array}{c}
    \gamma_{ \rm 0_{ \rm no ssc}} = \frac{3}{4} \left( \frac{\eta_{ \rm inj} v_{ \rm a}}{2 n_{0} R_{ \rm T} \sigma_{ \rm T} c (1 + u'_{ \rm ph}/u'_{ \rm B})} \right)^{\frac{1}{2}} \approx \frac{3}{4} \left( \frac{\eta_{ \rm inj}}{2 n_{0} R_{ \rm T} \sigma_{ \rm T} (1 + u'_{ \rm ph}/u'_{ \rm B})} \right)^{\frac{1}{2}} ,
    \\
    \tau_{ \rm c_{ \rm no ssc}} = \frac{9 m_{ \rm e} c^{2} }{16 \sigma_{ \rm T} c (u'_{ \rm B} + u'_{ \rm ph}) \gamma_{0}}.
    \end{array}
\end{equation}
In the limit that either synchrotron or EIC is the sole dominant radiation mechanism,
\begin{equation}
    \label{eq:gam0_EIC_syn_limit}
    \begin{array}{c}
    \gamma_{ \rm 0_{ \rm i}} = \frac{3}{4} \left( \frac{\eta_{ \rm inj} v_{ \rm a} u'_{ \rm B}}{2 n_{0} R_{ \rm T} \sigma_{ \rm T} c (u_{ \rm i})} \right)^{\frac{1}{2}} \approx \frac{3}{4} \left( \frac{\eta_{ \rm inj} u'_{ \rm B}}{2 n_{0} R_{ \rm T} \sigma_{ \rm T} (u_{ \rm i})} \right)^{\frac{1}{2}} ,
    \\
    \tau_{ \rm c_{ \rm i}} = \frac{9 m_{ \rm e} c^{2} }{16 \sigma_{ \rm T} c (u_{ \rm i}) \gamma_{0}},
    \end{array}
\end{equation}
where $u_{ \rm i}$ is $u'_{ \rm ph}$ if EIC is dominant or $u'_{ \rm B}$ if synchrotron is dominant. Using the synchrotron dominant case, the estimate for the synchrotron bolometric luminosity and synchrotron peak frequency are given below,
\begin{equation}
    \label{eq:syn_em_approx}
    \begin{array}{c}
    L_{ \rm bol} \approx \eta_{ \rm inj} L_{ \rm j 46} R_{ \rm tm}^{2}\Gamma_{ \rm j 1}^{2} 6.6 \times 10^{48} \: \mathrm{erg\,s}^{-1},
    \\
    \nu_{ \rm pk.syn} \approx \frac{\eta_{ \rm inj}\sigma \Gamma^{2}_{ \rm j1}}{R_{ \rm tm}L_{ \rm j46}}^{1/2} 3.6 \times 10^{12} \: \mathrm{Hz}.
    \end{array}
\end{equation}
Note that hereafter, we adopt the notation $Q = Q_{X} \times 10^{X}$ in CGS units. 

The radiative cooling of the particle distribution results in a observable emission signature.Synchrotron luminosity from the emission region is calculated from the emissivity $j'_{ \rm \nu}$, under the assumption that radiation is emitted isotropically \citep{Gould1979} while the blob moves directly in line with the observer i.e., the angle from the blobs motion to the line of sight of the observer $\theta_{ \rm obs} = 0$,

\begin{equation}
    \label{eq:nuLnu}
    \nu L_{ \rm \nu} = 3 \frac{f(\tau'_{ \rm \nu'})}{\tau'_{ \rm \nu'}} \delta^{4} (\frac{4\pi}{3}) R^{3}_{ \rm j} \nu' j'_{ \rm \nu'}, 
\end{equation}
where $\delta \equiv [\Gamma_{ \rm j}(1-\beta cos(\theta_{ \rm obs}))]^{-1}$ is the Doppler factor, $\tau_{ \rm \nu} \equiv 2 R_{ \rm T} \kappa_{ \rm \nu}$, $\kappa_{ \rm \nu}$ is the synchrotron self-absorption coefficient, and 
\begin{equation}
    \label{eq:opt_depth_function}
    f(\tau) = \frac{1}{2} + \frac{\exp(-\tau)}{\tau} - \frac{1 - \exp(-\tau)}{\tau^{2}},
\end{equation}
is the optical depth function \citep{Gould1979}. Compton luminosity is calculated similarly but since the only absorption mechanism we are incorporating is SSA, Compton luminosity takes the form of,
\begin{equation}
    \label{eq:nuLnu_c}
    \nu L_{ \rm \nu} =  \delta^{4} (\frac{4\pi}{3}) R^{3}_{ \rm j} \nu' j'_{ \rm \nu' c}, 
\end{equation}
where $j'_{ \rm \nu' c}$ is the Compton emissivity.

\subsection{Fitting Algorithm}
\label{sec:fitter}
One of the main advantages of this model is ability to test global parameters in a computationally efficient manor. To take full advantage of this, a fitting algorithm was developed in house to compare the model against other models or observations. This will not only be able to test the model but will allows to infer parameters about the physical systems.

The algorithm is a modified gradient descent algorithm. Instead of taking the gradient of the error each iteration, it instead takes the partial derivative of a given parameter for a set number of iterations before moving onto the next parameter,
\begin{equation}
    \label{eq:partial_derivative}
    g= \frac{Error[i] - Error[i-1]}{Parameter_j[i] - Parameter_j[i-1]}.
\end{equation}
Here g is the numerical partial derivative and $Parameter_j[i]$ denotes the value of a given parameter at the iteration i. The $Error[i]$ is a user defined error function that will compare the model to the 'true' data. For the work of this paper, we use a chi squared function \citep{arfken2013mathematical} to compare values from the 'true' data with values from the fit that have the closest x coordinate. We then use this derivative to indicate the new parameter for the next iteration,
\begin{equation}
    \label{eq:new_param}
    Parameter_j[i+1] = Parameter_j[i] - c_j g.
\end{equation}
$c_j$ is a constant multiple that slows down or speeds up the 'learning' process. It repeats this for a specified number of iteration and at the end shows the best fit. More details can be found at the GitHub page\footnote{\url{https://github.com/zkdavis/Base_ModelFitter.git}}.

\section{Results}
\label{sec:results}

In this section we explore predictions of our model and how they compare with observations. First, we observe the effects of large values of magnetization ($\sigma > 10$) on the underlying particle distribution. Then we test our models ability to reproduce a quiescent blazar SED. This is done by fitting our model to a dozen blazar SEDs found in \citet{Abdo2010}.

\subsection{Particle Distribution}

\begin{figure}
    \centering
    \includegraphics[width=0.48\textwidth]{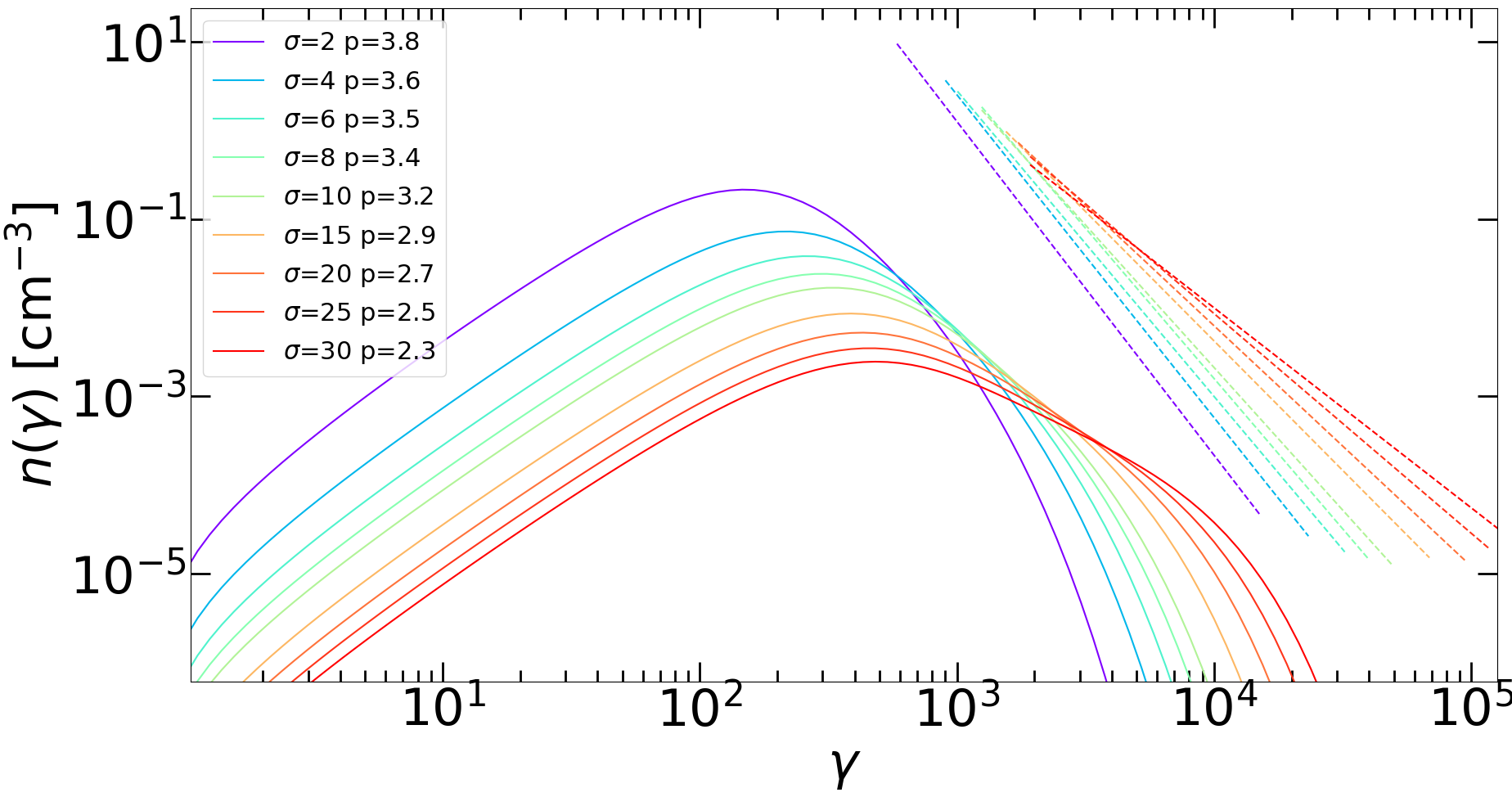}
    \caption{Steady states found for varying $\sigma$ from 1 to 30 with $\eta_{ \rm inj}=1$, $R_{ \rm m}=1$, $R_{ \rm TM}=0.4$, $L_{ \rm jm}=1$, $\Gamma=1$, and $\eta_{ \rm ph}=0.1$. The colors indicate the value of $\sigma$ with solid lines representing the particle distribution during the steady state and dashed lines are the power law slope for the hardened part of the spectrum.}
    \label{fig:particle_distribution}
\end{figure}

Turbulent acceleration is well modelled by resonant wave scattering or a second order Fermi process \citep{schlickeiser1989,Comisso2019,Demidem2020,Zhdankin2020}. With this comes the expectation for a hardened particle distribution above the thermal peak. The high energy part of the distribution can be described by  a power law functional dependence with an index $p$. To analyze the particle distribution created by turbulent acceleration and its dependence on plasma magnetization, we computed several distributions by holding all model parameters constant except for $\sigma$, allowing it to vary in the range $1\le \sigma \le30$. This parametric study shows a hardening of the power law spectra above the thermal peak for increasing magnetization.\footnote{The power laws are computed by averaging the slope from the peak to two standard deviations past the peak Lorentz factor,  i.e., it is averaged from $\gamma_{ \rm pk}$ to a Lorentz factor $\gamma_{ \rm 2}$ where $\gamma_{ \rm 2}$ satisfies the condition: n($\gamma_{ \rm 2}>\gamma_{ \rm pk}$) =$n(\gamma_{ \rm pk})$ - stdev($n(\gamma))*2$.} Shown in figure \ref{fig:particle_distribution}, we note that this turbulent model predicts hard tails for large values of $\sigma$. However, the tails only span a couple orders of magnitude in particle energy. As discussed in \citet{Zhdankin2020}, the steady state distribution is mostly thermal at low $\sigma$ but changes to include a non-thermal tail as $\Gamma_{ \rm a}$ increases with $\sigma$. One can note from equation \ref{eq:fpss} that if $\Gamma_{ \rm a}$ and $\Gamma_{ \rm h}$ are negligible, as they are for low $\sigma$, we get back a Maxwell–Jüttner distribution. Furthermore, from equation \ref{eq:fpss}, we can see the roles of the coefficients; $\Gamma_{ \rm h}$ helps setting the exponential cutoff and the fraction $\frac{\Gamma_{ \rm a}}{\Gamma_{ \rm 2}}$ almost completely describes the power law slope of the non-thermal particles. Where the particle power law $p\propto -\frac{\Gamma_{ \rm a}}{\Gamma_{ \rm 2}}$. However, this may break down for large $\sigma$ when $\Gamma_{ \rm a} > 0$ at which point it no longer describes an energy loss but rather a first order Fermi-like energy gain.

\subsection{Blazar Fits}
To test our model ability  to reproduce the quiescent emission of blazars, we use data from 12 sources from \citet{Abdo2010} which are representative populations of the total 48 quiescent SEDs reported in that work. The results of the blazar fits are best split into two categories. The first containing high synchrotron peaked (HSP) and intermediate synchrotron peaked (ISP) sources, while the second consist of the low synchrotron peaked (LSP) sources. Here LSP is defined as having a synchrotron peak frequency $\nu_{ \rm pk} \lesssim 10^{14} Hz$, ISP has a peak synchrotron frequency of $10^{14} \lesssim \nu_{ \rm pk} \lesssim 10^{15} Hz$, and HSP has $\nu_{ \rm pk} \gtrsim 10^{15} Hz$ \citep{Abdo2010}. All blazar SED and optical classifications are adopted from \citet{Abdo2010}. Further, for discussion purposes, we will group ISP and HSP into one category referred to as HSP for all sources with $\nu_{ \rm pk} \gtrsim 10^{14} Hz$.

To find the best fit for the 7 free parameter used in our model, we apply the model fitter described in section \ref{sec:fitter}. This algorithm requires an initial guess and bounds for the parameters. The closer the initial guess, and the bounds around the guess, the less iterations are needed to get a fit. However, the bounds are usually left very large so as to account for any unique possibilities. All of which can be seen in table \ref{tab:init_fit_parameters}. Parameters $\eta_{ \rm inj}, R_{ \rm TM}$, and $\eta_{ \rm ph}$ are all defined as a fraction of a whole so each of their max values are 1. The initial guess of 0.1 for each is based on the expectation that these will be a fraction and not an order of unity. The min of $10^{-4}$ for $\eta_{ \rm ph}$ is because we consider the case of negligible EIC cooling. The min of 0.01 for $\eta_{ \rm inj}$ is due to the fact that too little energy injected into the acceleration region would result in negligible emission. $R_{ \rm TM}$'s range is large so that its dependence in this model can be studied. As can be seen in synchrotron limit (equation \ref{eq:syn_em_approx}), $R_{ \rm TM}$ plays large role in dictating the peak frequency and bolometric luminosity. With the expectation that flaring events are caused by instabilities in the jet, and further that these instabilities later drive the turbulence, we expect the distance from the central engine to be related to the turbulence scales and thus related to the variability of the jet \citep{Bottcher2019},
\begin{equation}
    \label{eq:R_of_tv}
    R_{ \rm j} \approx \delta^{2} c t_{ \rm v_{ \rm obs}} \Gamma_{ \rm j} R_{ \rm TM}^{-1}(1+z)^{-1},
\end{equation}
where $t_{ \rm v_{ \rm obs}}$ is the observed variation time, and z is the redshift. For $\Gamma_{ \rm j} = 10$, $\theta_{ \rm obs}$ = 0,$R_{ \rm TM}$ = 1,$\Gamma_{ \rm j}=1$, z = 0, and $t_{ \rm v_{ \rm obs}}=$ 1 day from X-ray variability \citep{Wagner1995}, we get $R_{ \rm j} \approx 10^{17}$ cm. Thus, for an initial guess of $R_{ \rm j}$ we adopt similar values but left large bounds for model exploration. $\Gamma_{ \rm j}$ max constraints come from radio observations of $\Gamma_{ \rm j} \gtrsim 40$ being extremely rare \citep{Lister2016}. Here, we limit our analysis to relativistic turbulence and so, adopt $\sigma \ge 1$. The initial guess of $\sigma =3$ is comfortably in the relativistic plasma regime with the max of 30 to cover any extreme magnetization possibilities. $L_{ \rm j}$ range and initial guess is chosen to correspond with blazar luminosity range found in \citet{Ghisellini2017}.

For each SED we fit the data from \citet{Abdo2010} by allowing the fitter to iterate a 1000 times.

\begin{table}
\centering

\begin{tabular}{|l|l|l|l|}
\hline
Parameter    & Initial & Min       & Max      \\ \hline
$\eta_{ \rm inj}$ & 0.1     & 0.01      & 1        \\
$R_{ \rm j,18}$      & 0.1       & $10^{-3}$ & $10^{2}$ \\
$R_{ \rm TM}$     & 0.1     & $10^{-4}$ & 1        \\
$\eta_{ \rm ph}$  & 0.1     & $10^{-4}$ & 1        \\
$L_{ \rm j,46}$     & 1       & $10^{-3}$ & $10^{3}$ \\
$\Gamma$     & 10      & 1         & 50       \\
$\sigma$     & 3       & 1       & 30       \\ \hline
\end{tabular}

\caption{Initial fit parameter and ranges used in the fitting algorithm. After a successful run the we would then rerun by shrinking the max and min round the new best fit parameter. Here $R_j = R_{ \rm j,18} 10^{18} $ cm and $L_{ \rm j} = L_{ \rm j,46}10^{46}$ ergs/s.}
\label{tab:init_fit_parameters} 
\end{table}

At which point, they are reran with the resultant best fit parameter as the initial values and the max and min bounds are shrunk around these new values. This process repeats until there is no noticeable reduction in error for a maximum of 4000 iterations. The best fit parameters are found in tables \ref{tab:best_fit_sim_param}, \ref{tab:best_fit_phys_param}  and the best fit SEDs can be found in the appendices \ref{fig:lsp_seds},\ref{fig:hsp_seds}. 

\begin{table*}
\centering
\begin{tabular}{llllllllll}
\hline
Object       & Sed Type & Optical Type & $\eta_{ \rm inj}$ & $R_{ \rm j,18}$ & $R_{ \rm TM}$ & $\eta_{ \rm ph}$ & $L_{ \rm j,46}$ & $\Gamma$ & $\sigma$ \\
\hline
j0238.4+2855                 & LSP      & FSRQ         & 0.65         & 0.341    & 0.434    & 0.964       & 1        & 14.09    & 2.08     \\
j0137.1+4751                 & LSP      & FSRQ         & 0.767        & 0.288    & 1        & 0.581       & 0.317    & 10.67    & 4.15     \\
j1159.2+2912                 & LSP      & FSRQ         & 0.885        & 0.105    & 0.684    & 0.436       & 0.145    & 13.61    & 1.56     \\
j1256.1-0547                 & LSP      & FSRQ         & 0.212        & 0.998    & 0.3      & 0.612       & 3.295    & 17.57    & 3.82     \\
j0238.6+1636                 & LSP      & BL Lac       & 0.307        & 2.609    & 0.886    & 0.991       & 1.584    & 21.62    & 1.23     \\
j0855.4+2009                 & LSP      & BL Lac       & 0.201        & 0.412    & 1        & 0.162       & 0.245    & 12.7     & 4.56     \\
j1719.3+1746                 & LSP      & BL Lac       & 0.989        & 3.111    & 1.97E-03 & 0.731       & 127.544  & 30.39    & 1.07     \\
j1058.9+5629                 & ISP      & BL Lac       & 0.425        & 1.06E-03 & 0.481    & 0.179       & 7.98E-03 & 23.29    & 8.47     \\
j1221.7+2814                 & ISP      & BL Lac       & 0.323        & 0.047    & 9.51E-03 & 0.781       & 14.571   & 36.61    & 3.33     \\
j0449.7-4348                 & HSP      & BL Lac       & 0.284        & 0.018    & 0.124    & 0.198       & 1        & 21.92    & 14.38    \\
j2000.2+6506                 & HSP      & BL Lac       & 0.051        & 10.563   & 0.274    & 0.076       & 3.194    & 13.01    & 4.98     \\
j2158.8-3014                 & HSP      & BL Lac       & 0.082        & 9.67E-03 & 0.049    & 0.992       & 7.087    & 46.06    & 14.02  \\
\hline
\end{tabular}
\caption{Best fit parameters found after a maximum of 4000 iterations.}
\label{tab:best_fit_sim_param} 
\end{table*}

\begin{table*}
\centering
\begin{tabular}{llllllllll}
\hline
Object       & Sed Type & Optical Type & $L_{ \rm J}$  & $R_{ \rm j}$        & $R_T$    & $\gamma_{0}$ & uB      & $u_{ \rm ph}$ & $n_{0}$    \\
\hline
j0238.4+2855                 & LSP      & FSRQ         & 1.00E+46 & 3.41E+17 & 1.05E+16 & 112.11       & 0.915    & 1.708    & 585.46   \\
j0137.1+4751                 & LSP      & FSRQ         & 3.17E+45 & 2.88E+17 & 2.70E+16 & 230.69       & 0.407    & 0.175    & 130.53   \\
j1159.2+2912                 & LSP      & FSRQ         & 1.45E+45 & 1.05E+17 & 5.27E+15 & 149.68       & 1.403    & 1.370    & 1.19E+03 \\
j1256.1-0547                 & LSP      & FSRQ         & 3.30E+46 & 9.98E+17 & 1.70E+16 & 132.81       & 0.351    & 0.402    & 122.43   \\
j0238.6+1636                 & LSP      & BL Lac       & 1.58E+46 & 2.61E+18 & 1.07E+17 & 137.93       & 0.025    & 0.018    & 26.67    \\
j0855.4+2009                 & LSP      & BL Lac       & 2.45E+45 & 4.12E+17 & 3.25E+16 & 215.07       & 0.153    & 0.016    & 44.66    \\
j1719.3+1746                 & LSP      & BL Lac       & 1.28E+48 & 3.11E+18 & 2.01E+14 & 305.9        & 1.399    & 11.419   & 1.74E+03 \\
j1058.9+5629                 & ISP      & BL Lac       & 7.98E+43 & 1.06E+15 & 2.19E+13 & 239.26       & 755.002  & 34.282   & 1.19E+05 \\
j1221.7+2814                 & ISP      & BL Lac       & 1.46E+47 & 4.72E+16 & 1.23E+13 & 149.01       & 694.979  & 370.475  & 2.78E+05 \\
j0449.7-4348                 & HSP      & BL Lac       & 1.00E+46 & 1.79E+16 & 1.01E+14 & 179.44       & 330.33   & 33.642   & 3.06E+04 \\
j2000.2+6506                 & HSP      & BL Lac       & 3.19E+46 & 1.06E+19 & 2.23E+17 & 327.78       & 0.003    & 2.19E-5        & 0.81     \\
j2158.8-3014                 & HSP      & BL Lac       & 7.09E+46 & 9.67E+15 & 1.04E+13 & 61.7         & 8.04E+03 & 744.511  & 7.63E+05 \\
\hline
\end{tabular}

\caption{\label{tab:best_fit_phys_param}  Best fit results found after a maximum of 4000 iterations. All values are in CGS units. }
\end{table*}

\begin{figure}
    \centering
    \includegraphics[width=0.48\textwidth]{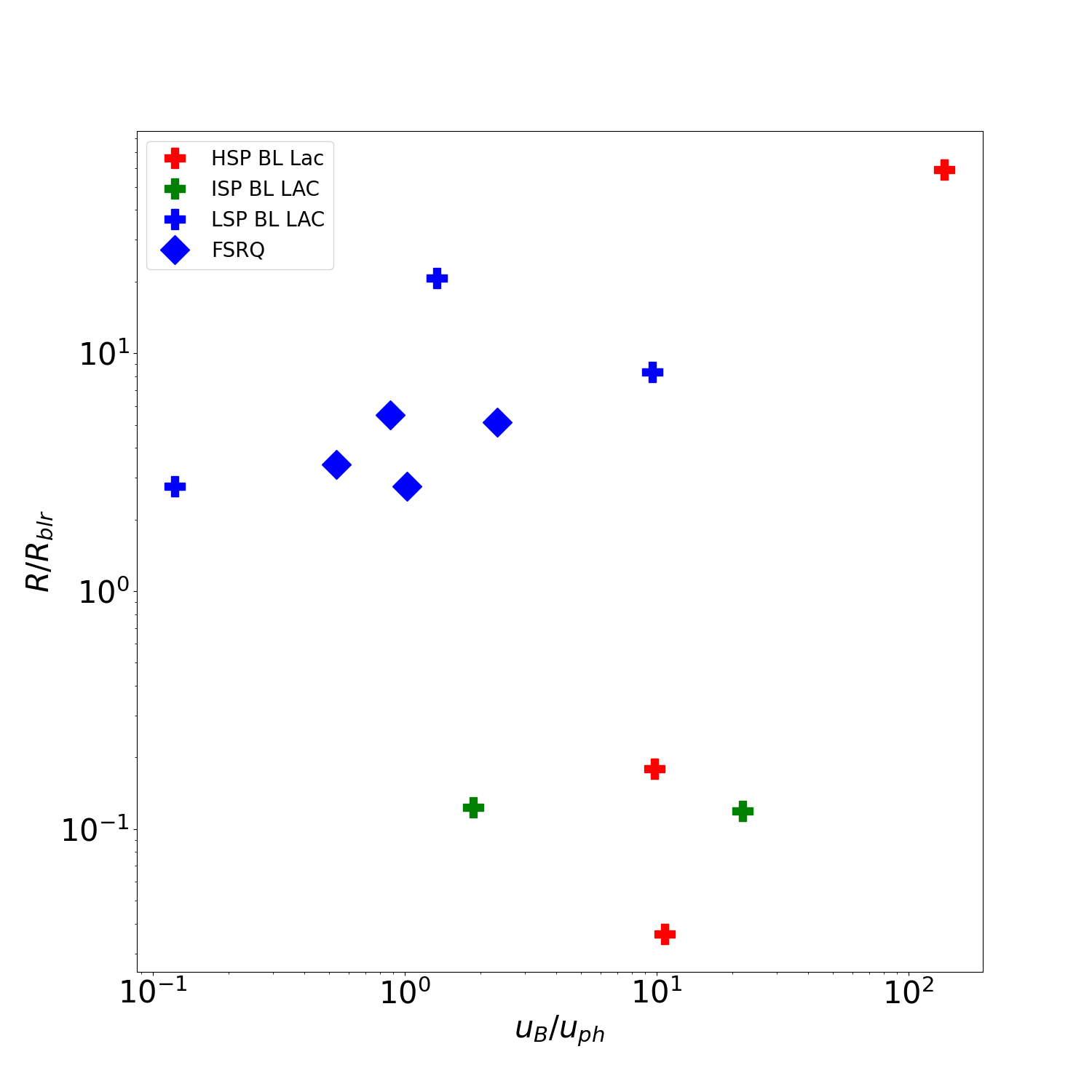}
    \caption{The y-axis displays how far a given source is from the central engine compared with the edge of the BLR region. In the x-axis we compare magnetic energy density to the BLR photon energy density.  }
    \label{fig:R_R_blr_ub_uph}
\end{figure}

\subsubsection{Low Synchrotron Peaked sources}
LSP sources make up 7 out of the 12 sources used in this paper and include BL Lacs as well as FSRQs. Resulting best fit SEDs for LSP sources can be seen in figures \ref{fig:lsp_seds}. These results are well described by our model. Generally, the fits exhibit a 3 peaked structure with the SSC and EIC working together to create what is generally the second peak in the typical double peaked structure \citep{Ghisellini2017}. In table \ref{tab:best_fit_sim_param}, we can see that these sources exhibit a range of $\Gamma_{ \rm j} \sim 10 - 30 $. A result consistent with most radio observation of LSP sources \citep{Lister2016}. Magnetization for these fits operates in a modest range of $\sigma \approx 1 - 4.5$. With $\sigma \approx 4$, according to figure \ref{fig:particle_distribution}, this would indicate a rather modest power law index of about 3.6 for the underlying particle distribution. For the majority of LSP sources, we infer a turbulent region that is a fraction of the jets cross section with $R_{ \rm TM} \lesssim 0.5$ while the rest are close to 1. When comparing with \citet{Ghisellini2017}, our LSP sources demonstrate a jet luminosity comparable to most blazars with $L_{ \rm j}$ between $10^{45}$ and $ 10^{46}$ ergs/s.\footnote{An exception is made here for j1719.3+1746. Though it is a LSP according to \citet{Abdo2010}, our fits would better describe this as an HSP source}  The emission region for most of these sources is just outside of the the broad line region with the magnetic energy density staying comparable to broad line regions photon field energy density (see figure \ref{fig:R_R_blr_ub_uph}). Still all emission regions remain within a parsec from the central engine. The mean particle Lorentz factor range is $\gamma_{0} \approx 100-200 $.

\subsubsection{High synchrotron peaked source}
\begin{figure}
    \centering
    \includegraphics[width=0.48\textwidth]{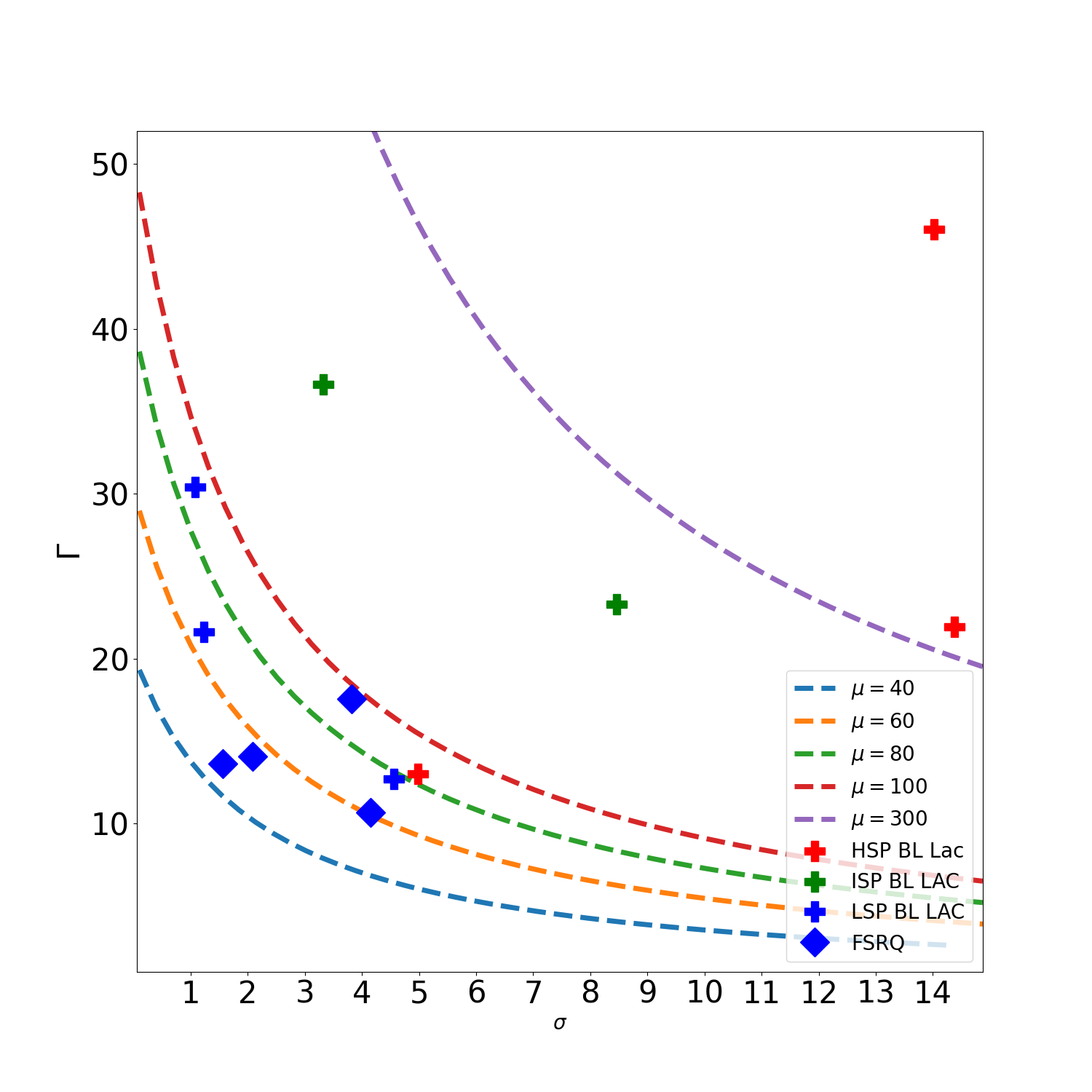}
    \caption{
Above we plot a given sources bulk Lorentz factor vs its magnetization. The dashed lines are baryon loading contours for values $\mu = 40,60,80,100,300$.
    }
    \label{fig:sigmavsgamma_baryoncontour}
\end{figure}

Contrary to the LSP fits, almost all of the HSP sources display a double peaked spectra \ref{fig:hsp_seds}. Where the second peak is usually much broader and dominated by SSC emission. The model accurately describes the compton peak but doesn't seem to be able to create as broad a synchrotron peak. HSP best fit parameters are much more extreme than that of the LSP\footnote{It should be noted that of this group, j2000.2+6506 is a clear out liar in our results. Though \citet{Abdo2010} initially categorized this as an HSP our fit is that of an LSP and thus the resulting fit being much closer to that of LSP sources is due to phenomenological fitting.}. This is perhaps best shown at figure \ref{fig:sigmavsgamma_baryoncontour}, where one notices a large jump in baryon number $\mu$ separating the HSP sources from the LSP sources. Here, we define the jet baryon number as $\mu=\Gamma_{ \rm j} ( 1 + \sigma)$, which also corresponds to the asymptotic bulk Lorentz factor of the jet provided that all the magnetic energy were to be converted into bulk motion.  From the same figure \ref{fig:sigmavsgamma_baryoncontour}, we can see $\Gamma_{ \rm j}$ ranges from 10-50. The large $\Gamma_{ \rm j}$ was not expected since the sources are all BL Lac objects and don't typically exhibit large Lorentz boosted compton peak like that of FSRQs \citep{Ghisellini2017}. As one might expect to have more energetic particles to account for the higher energy of emission in HSPs, the magnetization for these sources is much higher than the LSP sources, with $\sigma_{ \rm ave} \approx 9$. This would also explain the higher magnetic energy density to photon field energy density seen in these sources in figure \ref{fig:R_R_blr_ub_uph}. The jet luminosity of these sources varies from $10^{44} - 10^{47}$erg/s. Emission regions for these sources are much closer to the central engine with $R_{ \rm j} \approx 10^{15} - 10^{16}$ cm. Following this, we have an even more compact turbulent region from and $R_{ \rm TM}$ about a 10 times smaller than that of the LSPs. Particle number density for the sources also seems to be much larger than the LSP sources with some having, a possibly problematic, $n_{0} \approx 10^{5}$ $cm^{-3}$. This is probably needed to create the large SSC emission that dominates these SEDs. With them being much closer to the central engine, these sources all have the emission region well within the broad line region (see figure \ref{fig:R_R_blr_ub_uph}).

\section{Discussion}
\label{sec:discussion}

Our model of turbulent acceleration has a unique advantage in that the particle distribution is not picked by hand but, rather, it arises from the physical properties of the large-scale emission region. Typically, modeling of the particle distribution inside relativistic jets is done by assuming the particles form a power law with an index $p$ within a range $\gamma_{ \rm min}$ through $\gamma_{ \rm max}$, or by more complicated particle distributions. Quantities such as  $p$, $\gamma_{ \rm min}$, $\gamma_{ \rm max}$ are usually treated as free parameters. This non-thermal particle distribution is then injected into a region where it cools radiatively while slowly escaping the emission region. In works such as \citet{Bottcher2013}, a steady state can be reached by balancing the injection of non-thermal particles with radiative cooling and particle escape. Fits from \citet{Bottcher2013}  are able to constrain the Doppler factor and can be compared to ours. For the source j12561.1-0547 (3C279), \citet{Bottcher2013} finds $\delta = 17$. Though $\delta$ is highly dependent on the observer angle, for our assumption of $\theta_{ \rm obs} = 0$ we found a $\delta \approx 35$. Putting our result within the bounds of superluminal studies such as \citet{Bloom2013} where the, admittedly broad, range is $\delta \approx 20-80$.

This work considers a scenario where turbulence is generated by the onset of an instability within the jet. The resulting turbulence energizes leptons which, at the same time experience radiative losses. As a result of a balance of energization and cooling, the particles in the turbulent plasma acquire a steady state distribution which has a distinctly non-thermal appearance. The emission of the leptons may be of relevance to the observed blazar emission and, in particular to the quiescent emission seen in these sources. As can be seen in figure \ref{fig:particle_distribution}, the heating/cooling balance in the turbulent region results in a narrow particle distribution for modest particle magnetization $\sim 1$ while, for sufficient high magnetization, the distribution broadens. In the case of high magnetization, we find extended particle distribution that can be approximated by a power-law for a few orders of magnitude in energy above the peak of the distribution. We can see in the SEDs found in appendix \ref{fig:lsp_seds},\ref{fig:hsp_seds} that this translates into smooth emission spectra that well describe the quiescent emission observed in blazars. Aside from the quiescent emission, of focus in this work, this model may have implications for blazar variability. In \citet{Marscher2014} turbulent plasma is used to account for observations of rather rapid variability in radiative flux and polarization seen multi-wavelength blazar observations by simulating turbulence with a large set of plasma cells that have a randomly oriented, but otherwise smooth, magnetic field. Here the collective effect of these cells should be similar to the turbulent region in this work.

The best-fit values for the jet magnetization and bulk Lorentz factor inferred from the model show consistency with the model put forward in \citet{Rueda_Becerril_2020}, where the majority of the sources appear to be launched with a similar baryon loading parameter $\mu \approx 40-100$. In figure \ref{fig:sigmavsgamma_baryoncontour} this shown with the noted exception of the HSP sources. HSPs appear to require significantly larger magnetization  $\mu \approx 350$ from other blazar sources. Although the model doesn't fit the spectrum of the HSP sources well, there may still be some information to glean from comparison. We find that the model generally favors a dense emission region for HSPs that is close to the central engine. We can gather an understanding for how the fits reach different peaks in luminosity and frequency by looking at the synchrotron limit equation \ref{eq:syn_em_approx}. In order to get to these high synchrotron frequencies, the turbulent scale needs to drop. However, the smaller the emission region, the dimmer the source. Though most of the HSP sources tend to be dimmer, the squared dependence of the luminosity with $R_{ \rm TM}$ is a steep one. The fitted model parameters may turn out to be less extreme if one includes Klein-Nishina (KN) correction to Compton scattering, neglected in this study. The HSP sources are dominated by SSC cooling and the reduction of the SSC cooling efficiency because of relativistic corrections would allow the particles to reach a higher $\gamma_0$ and broaden the synchrotron emission for the same parameters. The KN corrections will tend to result to smaller values for $\mu$  for HSP sources, since the drop in cooling efficiency provided by the KN cross section may drop the energy  per baryon required to reach such high synchrotron peaks.
It may also be the case that the extreme fit parameters simply suggest that HSP sources require more efficient acceleration than the turbulence prescription adopted here can provide.

\subsection{Future Extensions to the Model}

Currently, there are few PIC simulations that have studied relativistic turbulence. This work is based on the PIC simulations in \citet{Zhdankin2020}, which are only a handful of simulations are reported that cover a small range of plasma magnetization. One could use Alfvén wave scattering to describe the diffusion in the plasma. Simulations have shown this to be a consistent description \citep{Comisso2019} but without a description for the advection coefficient our only means to improve our diffusive model is to include more PIC simulation data at additional magnetizations.

For this work, we develop a simple single-zone model for the blazar region but extending this model to a include multiple emission zones should be straightforward. The easiest approach might be one similar to that found in \citet{boula2021twozone} where the cooling and heating take place in distinct regions. One can imagine having the steady state reached before exiting an acceleration region. The acceleration region could have minimum photon cooling and create a steady state by balancing turbulent heating with the synchrotron losses alone. Once exiting the acceleration region, the particles would then be exposed to photon fields like those of the BLR. This would effectively separate the parameters of the jet that create the synchrotron peak from those that create the compton peak. This two zone model would still use a particle distribution that arises from the global parameters of the emission region, reducing the assumption about the particle acceleration, but would also allows us to compute the synchrotron and Compton emission with model parameters that are independent of each other.

Another natural extension to this model is to account for the time-dependent nature of blazar emission. For instance, one may recreate and build on a scenario similar to the minijet model proposed in \citet{Giannios2009}. Here, instead of a single turbulent region to represent the total acceleration region, one can envision several compact turbulent regions each resulting in non-thermal particle acceleration that contributes to the total emission. This picture may be particularly applicable to HSP sources with the smaller inferred values $R_{ \rm TM}$ possibly representing a compact turbulent regions within the jet. Such regions may be driven by plasma outflowing from large-scale reconnecting current sheets. The fast-evolving flares originate in the current sheets \citep{2019Christie} while slower evolving and quiescent emission is result of the turbulent heating discussed here.

\section{Conclusion}
\label{sec:conclusion}
In this paper, we have constructed a turbulent jet model that incorporates the plasma physics from PIC simulations while being computationally efficient enough to run a sufficient number of models for the jet emission to constrain the physical properties of blazars. 
By analyzing work done by \citet{Zhdankin2020}, we found diffusion and advection coefficients that can describe a particle distribution at much a larger scales and magnetizations than PIC simulations have available, while retaining the crucial microscopic physics from the simulations. 
We then used this information to build a single zone turbulent jet model where the emission comes from a blob of plasma in which particles are accelerated by turbulence and cooled radiatively. Since, turbulence may not be a fast enough accelerator to account for, say, fast evolving, intense  flaring blazar events, we focused on the model's ability to recreate the observed quiescent emission. Using data from \citet{Abdo2010}, we compared our model predictions against the broad band SEDs of 12 blazars. We did this by performing a fit of the 7 free parameters in the model over a large parameter domain. These 7 parameters are key insights into the blazar's emission region. For LSP sources we found that the emission region is typically at the, or slightly beyond, the edge of the BLR region. The emission region's size itself is typically a modest fraction of a cross section of the jet but could be order of unity. Perhaps hinting at distinct instabilities that trigger turbulence in different sources. The magnetization inferred by the model suggests that LSP sources are moderately Poynting dominated with $\sigma = 1-5$ with a bulk Lorentz factor in the range of 10-30. Contrary to the LSP sources, the HSP and ISP sources are generally less satisfactory fits that require more extreme parameters. HSPs and ISPs require dense emission regions that are closer to the central engine. These jets are inferred to be largely Poynting flux dominated with $\sigma = 5-15$ and have extreme bulk Lorentz factors up to 50. While the HSP and ISP fits are extreme, our model accurately reproduces quiescent emission from LSP blazars. It is able to do this with a successful turbulent model that accurately encodes expensive results from PIC simulations into a computationally efficient kinetic description of the turbulent plasma. 

\section*{Acknowledgements}

We would like to thank Vladimir Zhdankin for providing additional PIC results that helped improve our physical modeling, and for his insightful comments. We are very grateful for the insightful comments offered by Maria Petropoulou. Z.D. and D.G. acknowledge support from the Fermi Cycle 14 Guest Investigator Program 80NSSC21K1951, 80NSSC21K1938 and the NSF AST-2107806 grants. J.M.R.B. acknowledge support from the NSF AST-2009330 grant.




\bibliographystyle{mnras}
\bibliography{example} 



\appendix



\section{Steady state SSC cooling in stochastic magnetic fields}
\label{sec:ssc_deriv}
Starting with power radiated via synchrotron \citep{Rybicki:1979} (eq. 6.33):
\begin{equation}
    \label{eq:power_syn}
    p_{ \rm s}(\nu) = \frac{\sqrt{3}e^{3}B\sin{\theta}\nu}{mc^{2} \nu_{ \rm c}}F\left(\frac{\nu}{\nu_{ \rm c}}\right) \quad \mathrm{erg\,s^{-1}\,Hz^{-1}}, 
\end{equation}
\begin{equation}
    \label{eq:nuc}
    \nu_{ \rm c}=\frac{3\gamma^{2} e B\sin{\theta}}{4\pi mc} = \frac{3}{2}\nu_{0} \gamma^{2}.
\end{equation}
Here, $\theta$ is the pitch angle of the particle, $\gamma$ is the particles lorentz factor and F is,
\begin{equation}
    \label{eq:syndistributionfunction}
    F(x)=\int^{\infty}_{ \rm \frac{x}{\sin{\theta}}} K_{ \rm 5/3}(z)dz.
\end{equation}
Combining \ref{eq:power_syn} and \ref{eq:nuc} we get:
\begin{equation}
    \label{eq:powersyn2}
    p_{\rm s}=\frac{4\pi e^2 \nu}{\sqrt{3}c\gamma^2}F\left(\frac{\nu}{\nu_{ \rm c}}\right)  \quad \mathrm{erg\,s^{-1}\,Hz^{-1}}.
\end{equation}
In a turbulent media the synchrotron power needs to be averaged over scattering angles. Here we follow \citet{CRISIUS1988},
\begin{equation}
    \label{eq:isopsint}
     p_{ \rm rs}=\frac{q_{0}\nu}{\gamma^{2}}\int^{2\pi}_{0} d\phi \int^{\pi}_{0} d\theta \; \sin{\theta} \int^\infty_{ \rm \frac{x}{\sin{\theta}}} dz \;  K_{ \rm 5/3}(z) \quad \mathrm{erg\,s^{-1}\,Hz^{-1}}.
\end{equation}
In \citet{CRUSIUS1986} it was shown that,
\begin{equation}
    \label{eq:csint}
     \int^{\pi}_{0} d\theta \; \sin{\theta} \int^\infty_{ \rm \frac{x}{\sin{\theta}}} dz \;  K_{ \rm 5/3}(z) = \pi CS(x),
\end{equation}
where,
\begin{equation}
    \label{eq:csofx}
    CS(x) = W_{ \rm 0,\frac{4}{3}}(x)W_{ \rm 0,\frac{1}{3}}(x) - W_{ \rm \frac{1}{2},\frac{5}{6}}(x)W_{ \rm \frac{-1}{2},\frac{5}{6}}(x),
\end{equation}
and $W_{ \rm i,j}$ denotes the Whittaker's function. Giving us an expression for the emitted power via synchrotron in a stochastic magnetic field \ref{eq:psyn_stoch}.
\begin{equation}
    \label{eq:psyn_stoch}
     p_{ \rm rs}(\nu)=\frac{q_{0}\nu}{\gamma^{2}}\frac{\pi}{2}CS(\frac{\nu}{\nu_{ \rm c}}) \qquad \mathrm{erg\,s^{-1}\,Hz^{-1}}
\end{equation}

It's easier to to keep up with Schlickeiser's derivation \citep{Schlickeiser2009} if we switch power from frequency dependence to energy dependence:
\begin{align}
    \label{eq:freq2energ}
    p(\nu)=\frac{d\epsilon}{d\nu dt}  \quad \mathrm{erg\,s^{-1}\,Hz^{-1}} \xrightarrow{}  
    p(\epsilon)=\frac{d\epsilon}{d\epsilon dt} = \frac{1}{h}p(\nu=\frac{\epsilon}{h}) \quad \mathrm{s^{-1}}.
\end{align}
This turns equation \ref{eq:psyn_stoch} into,
\begin{equation}
    \label{eq:psyn_stoch_ofE}
     p_{ \rm rs}(\epsilon)=\frac{q_{0}\epsilon}{\gamma^{2} h^{2}}\frac{\pi}{2}CS(\frac{\epsilon}{\epsilon_{ \rm c}}) \qquad \mathrm{s^{-1}}.
\end{equation}
Now lets take a look at what the power emitted via SSC should be \citet{Schlickeiser2009} eq.4.2,
\begin{equation}
    \label{eq:pssc0}
    p_{ \rm ssc}(\epsilon_{ \rm \gamma},\gamma) = c\epsilon_{ \rm \gamma}\int^{\infty}_{0}d\epsilon n_{ \rm s}(\epsilon) \sigma(\epsilon_{ \rm \gamma},\epsilon,\gamma) \qquad \mathrm{s^{-1}}.
\end{equation}
The total power can be found by integrating over the scattered energies,
\begin{equation}
    \label{eq:pssc1}
    P_{ \rm ssc}(\epsilon_{ \rm \gamma},\gamma) = c\int^{\infty}_{0}d\epsilon_{ \rm \gamma}\epsilon_{ \rm \gamma}\int^{\infty}_{0}d\epsilon n_{ \rm s}(\epsilon) \sigma(\epsilon_{ \rm \gamma},\epsilon,\gamma) \qquad \mathrm{erg\,s^{-1}}.
\end{equation}
Here, $n_{ \rm s}$ is the number density of scattered photons, $\epsilon_{ \rm \gamma}$ is the scattered photon energy, $\gamma$ is the electron's Lorentz factor, $\sigma$ is the interaction cross section, and $\epsilon$ is the photons prescattered energy. The cross section is given by \citet{Schlickeiser2009} eq. 4.2.1:
\begin{equation}
    \label{eq:sigma0}
    \sigma(\epsilon_{ \rm \gamma},\epsilon,\gamma) = \frac{3\sigma_{ \rm T}}{4\epsilon\gamma^{2}}G(\epsilon,\Gamma) \qquad \mathrm{cm^{2}\,erg^{-1}}.
\end{equation}
The function G is beyond the scope of this derivation but accounts for Klein-Nishina affects and is described in \citet{Schlickeiser2009}. The parameter $\Gamma = \frac{4\epsilon\gamma}{mc^{2}}$. Plugging in \ref{eq:sigma0} and making the substitution $q = \frac{\epsilon}{\Gamma(\gamma m c^{2} - \epsilon_{ \rm \gamma}}$, we are able to turn \ref{eq:pssc1} into,
\begin{equation}
    \label{eq:pssc2}
    P_{ \rm ssc}(\epsilon_{ \rm \gamma},\gamma) = \frac{3c\sigma_{ \rm T}}{4\gamma^{2}} \int^{\infty}_{0} d\epsilon \frac{n_{ \rm s}(\epsilon,t)}{\epsilon}\int^{\infty}_{0} d\epsilon_{ \rm \gamma} \epsilon_{ \rm \gamma} G(q,\Gamma)  \quad \mathrm{erg\,s^{-1}}.
\end{equation}
Here we are going to focus on the sub KN regime($\Gamma < 1$ where ssc has the strongest cooling and assume that $\Gamma > 1$ has a negligible effect. More precisely, $G(q,\Gamma > 1) = 0$. This also puts a limit on $\epsilon \leq \frac{mc^{2}}{4\gamma}$. $\epsilon_{ \rm \gamma}$ is limited by the amount of energy it gain in a head on collision $\epsilon_{ \rm \gamma} \leq \frac{\Gamma \gamma mc^{2}}{\Gamma + 1} $. Applying this and changing integration variable to q we get,
\begin{equation}
    \label{eq:pssc3}
    P_{ \rm ssc}(\gamma,t) = 12 \sigma_{ \rm T} c \gamma^{2} \int^{\frac{mc^{2}}{4\gamma}}_{\rm 0} d\epsilon \epsilon  n_{ \rm s}(\epsilon,t)\int^{1}_{0} dq \frac{qG(q,\Gamma)}{(1 + \Gamma q)^{3}} \quad \mathrm{erg\,s^{-1}}.
\end{equation}
Here the right most integral can be approximated as,
\begin{equation}
    \label{eq:crosssectionint}
    \int^{1}_{0} dq \frac{qG(q,\Gamma)}{(1 + \Gamma q)^{3}}  \simeq \frac{1}{9} \quad for \quad \Gamma << 1. 
\end{equation}
Combining this result with eq \ref{eq:pssc3},
\begin{equation}
    \label{eq:pssc4}
    P_{ \rm ssc}(\gamma,t) = \frac{4}{3} \sigma_{ \rm T} c \gamma^{2} \int^{\frac{mc^{2}}{4\gamma}}_{0} d\epsilon \epsilon  n_{ \rm s}(\epsilon,t)  \qquad \mathrm{erg\,s^{-1}}.
\end{equation}
The synchrotron photon density spectrum is given by,
\begin{equation}
    \label{eq:numofsynchphotons}
    n_{ \rm s}(\epsilon,t) = \frac{4\pi R_{ \rm em}}{c \epsilon}j_{ \rm s}(\epsilon,t) \qquad \mathrm{cm^{-3}\,erg^{-1}},
\end{equation}
where $R_{ \rm em}$ is the size of the emission region and $j_{ \rm s}(\epsilon,t)$ is given by,
\begin{equation}
    \label{eq:synem}
    j_{ \rm s}(\epsilon,t) = \frac{1}{4\pi}\int^{\infty}_{0} d\gamma n(\gamma,t)p_{ \rm rs}(\gamma,t) \qquad \mathrm{cm^{-3}}.
\end{equation}
Combining \ref{eq:numofsynchphotons} with \ref{eq:synem} we get,
\begin{equation}
 \label{eq:numossynch1}
  n_{ \rm s} = \frac{R_{ \rm em}}{c\epsilon} \int^{\infty}_{0} d\gamma n(\gamma,t)\frac{q_{0} \epsilon}{2h^{2} \gamma^{2}} \pi CS(x)  \qquad \mathrm{cm^{-3}erg^{-1}}.
\end{equation}
Where $x=\frac{2\epsilon}{3\epsilon_{0}\gamma^{2}}$. Combining \ref{eq:numossynch1} with \ref{eq:pssc4}, and changing the integration variable to $x$ we obtain,
\begin{align}
    P_{\rm ssc} = & \frac{3 \pi \sigma_{ \rm T} c \gamma^{2} q_{0} R_{ \rm em} \epsilon_{0}^{2}}{2 h^{2} c} \int^{\infty}_{0} d\gamma \gamma^{2} n(\gamma,t) \, \times \nonumber \\
    & \int^{\frac{m c^{2}}{6 \gamma^{3} \epsilon_{0}}}_{0} dx x CS(x) \quad \mathrm{erg\,s^{-1}}. \label{eq:pssc5}
\end{align}
To find a nice analytic solution we limit ourselves to regime where $CS(x)$ is dominant. This happens when $\gamma \leq \sqrt{\frac{mc^{2}}{6\epsilon}}$. So here we assume $CS(x) = 0$ for $\gamma>\gamma_{ \rm KN} = \sqrt{\frac{mc^{2}}{6\epsilon_{0}\gamma}}$. With these assumptions equation \ref{eq:pssc5} becomes,
\begin{align}
    P_{\rm ssc} = & \frac{3\pi \sigma_{ \rm T} c \gamma^{2} q_{0} R_{ \rm em} \epsilon_{0}^{2}}{2 h^{2} c} \int^{\infty}_{0} d\gamma \gamma^{2} n(\gamma, t) \,\times \nonumber\\
     & \int^{\frac{\gamma_{ \rm KN}}{\gamma}}_{0} dx x CS(x) \qquad \mathrm{erg\,s^{-1}}. \label{eq:pssc6}
\end{align}

\ref{eq:pssc6} can equivalently be described by,
\begin{align}
    P_{\rm ssc} = & \frac{3\pi \sigma_{ \rm T} c \gamma^{2} q_{0} R_{ \rm em} \epsilon_{0}^{2}}{2 h^{2} c} \int^{\gamma_{ \rm KN}}_{0} d\gamma \gamma^{2} n(\gamma,t) \,\times \nonumber \\
    & \int^{\infty}_{0} dx x CS(x)  \qquad \mathrm{erg\,s^{-1}}, \label{eq:pssc7}
\end{align}

where the x dependent integral is now given by \citet{Schlickeiser2009} eq 16,
\begin{equation}
    \label{eq:c1}
    c_{ \rm 1} = \int^{\infty}_{0} dx x CS(x)  = \frac{32}{81}\sqrt{3} .
\end{equation}
Finally, if we limit are distribution to $\gamma<\gamma_{ \rm KN}\simeq 1.94\times 10^{4} B^{=1/3}$ we can extend the limit to infinity. Resulting in,
\begin{equation}
    \label{eq:pssc7}
    P_{\rm ssc} = \frac{3\pi \sigma_{ \rm T} c \gamma^{2} q_{0} R_{ \rm em} \epsilon_{0}^{2} c_{ \rm 1}}{2 h^{2} c} \int^{\infty}_{0} d\gamma \gamma^{2} n(\gamma,t) \qquad \mathrm{erg\,s^{-1}},
\end{equation}
noting that,
\begin{equation}
    \label{eq:expectationofgamma}
    \int^{\infty}_{0} d\gamma \gamma^{2} n(\gamma,t) = \left<\gamma^{2}(t)\right> n_{0}
\end{equation}
we get our final expression,
\begin{equation}
    \label{eq:gamdot}
    \dot{\gamma}_{ \rm ssc} = \frac{3\pi \sigma_{ \rm T} c q_{0} R_{ \rm em} \epsilon_{0}^{2} c_{ \rm 1} n_{0} \gamma^{2} \left<\gamma^{2}(t)\right>}{2 h^{2} c m c^{2}}  \qquad \mathrm{s^{-1}}.
\end{equation}
To simplify remember that $q_{0} = \frac{4\pi e^{2}}{\sqrt{3}c}$, $\epsilon = \frac{eBh}{2\pi mc}$, $u_{ \rm B} = \frac{B^{2}}{8\pi}$. Defining $P_{0} = \frac{e^2}{\hbar^{2} c 2 \sqrt{3}}$ and $A_{0}$ as,
\begin{equation}
    \label{eq:a0}
    A_{0} =\frac{ 3 c_{ \rm 1} \sigma_{ \rm T} P_{0} \hbar^{2} e^{2} 8 \pi}{m^{3}c^{4}} ,
\end{equation}
we can find $\dot{\gamma}_{ \rm ssc}$ to be \ref{eq:gdotfinal},
\begin{equation}
    \label{eq:gdotfinal}
    \dot{\gamma}_{ \rm ssc} = A_{0} R_{ \rm em} n_{0} u_{ \rm B} \gamma^{2} \left<\gamma^{2}(t)\right> \qquad \mathrm{s^{-1}} .
\end{equation}

\clearpage

\section{LSP SEDs}

\noindent\begin{minipage}{\textwidth}
    \centering
    \includegraphics[height = 0.9\textheight]{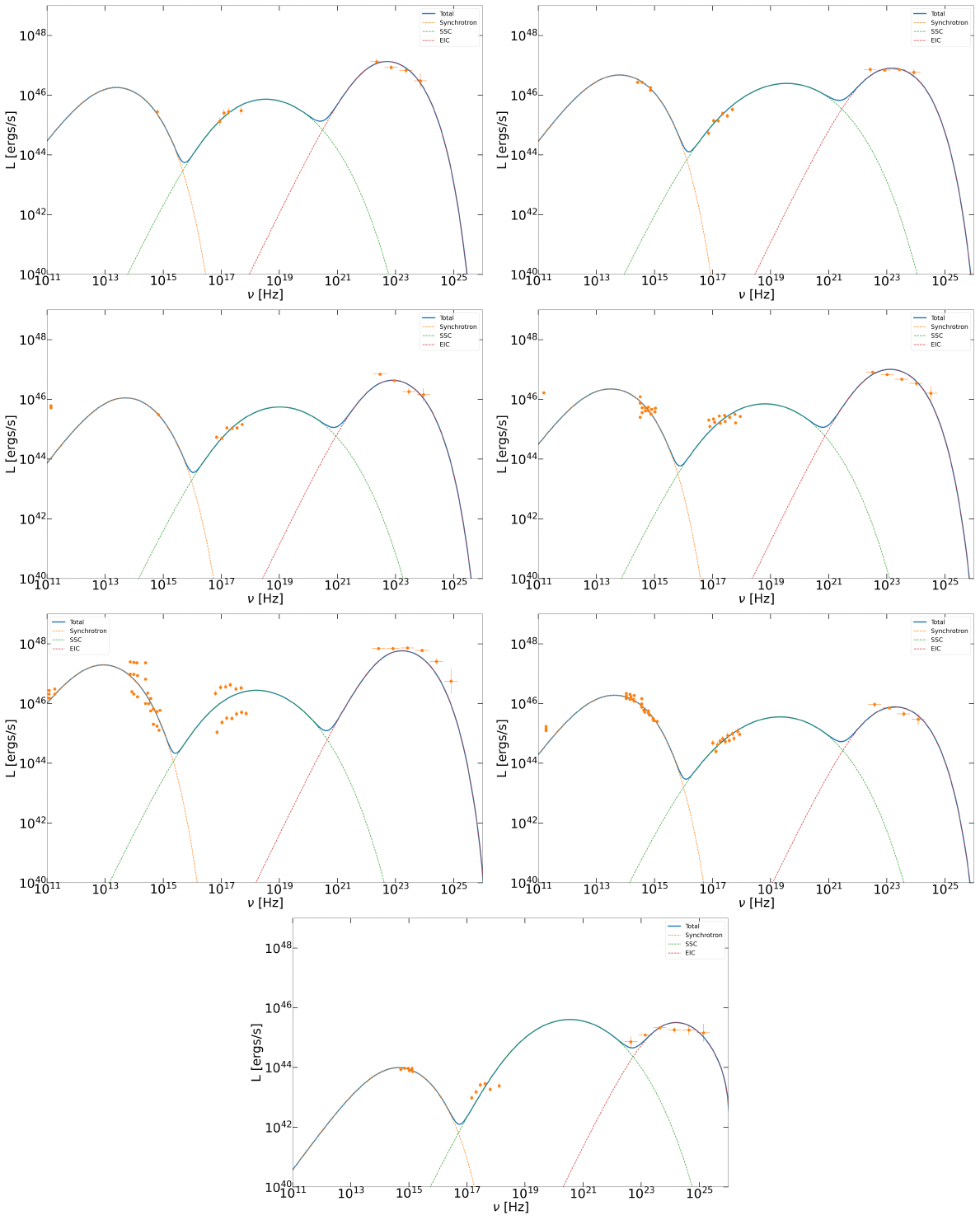}
    \captionof{figure}{Resultant best fit LSP SEDs. From left to right the sources are: j0238.4+2855, j0137.1+4751, j1159.2+2912, j1256.1-0547, j0238.6+1636, j0855.4+2009 and j1719.3+1746.}
    \label{fig:lsp_seds}
\end{minipage}

\clearpage
\section{HSP SEDs}

\noindent\begin{minipage}{\textwidth}
    \centering
    \includegraphics[width = 1\textwidth]{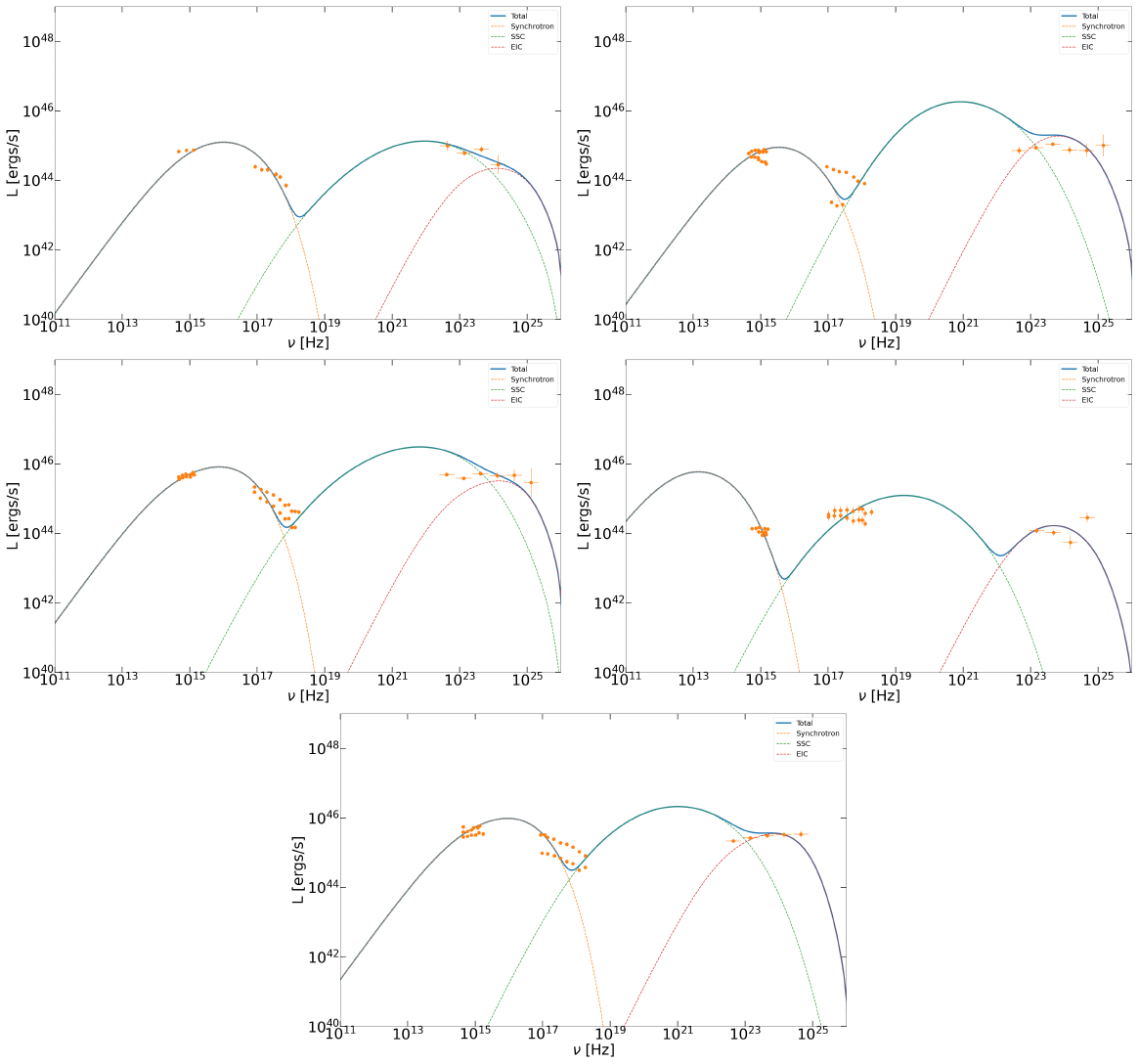}
    \captionof{figure}{Resultant best fit HSP and ISP SEDs. From left to right the sources are: j1058.9+5629, j1221.7+2814, j0449.7-4348, j2000.2+6506 and j2158.8-3014.}
    \label{fig:hsp_seds}
\end{minipage}

\end{document}